\def\opalackerstaff{OPAL Collaboration, K.\ Ackerstaff \etal}
\def\opalabbiendi{OPAL Collaboration, G.\ Abbiendi \etal}

\def\opalahmet{OPAL Collaboration, K.\ Ahmet \etal}
\newcommand{\PLB}[3]  {Phys.\ Lett.\ \textbf{B#1} (#2) #3}
\newcommand{\ZPC}[3]  {Z.\ Phys.\ \textbf{C#1} (#2) #3}
\newcommand{\EPC}[3]  {Eur.\ Phys.\ J.\ \textbf{C#1} (#2) #3}
\newcommand{\NIMA}[3] {Nucl.\ Instr.\ and Meth.\ \textbf{A#1} (#2) #3}

\newcommand{\PRL}[3]  {Phys.\ Rev.\ Lett.\ \textbf{#1} (#2) #3}
\newcommand{\PRD}[3]  {Phys.\ Rev.\ \textbf{D#1} (#2) #3}
\newcommand{\NPB}[3]  {Nucl.\ Phys.\ \textbf{B#1} (#2) #3}

\newcommand{\CPC}[3]  {Comp.\ Phys.\ Comm.\ \textbf{#1} (#2) #3}
\newcommand{\epem}{\mbox{$\mathrm{e^+e^-}$}}
\newcommand{\epmem}{\mbox{$\mathrm{e^\pm e^-}$}}

\newcommand{\Zzero}{\mbox{${\mathrm{Z}^0}$}}
\newcommand{\Zgamma}{\mbox{${\mathrm{Z}^0} / \gamma$}}
\newcommand{\WW} {\mbox{$\mathrm{W^+W^-}$}}

\newcommand{\Wmv} {\mbox{$W^{\mu\nu}$}}

\newcommand{\Wma} {\mbox{$W_{\mu\alpha}$}}
\newcommand{\Wka} {\mbox{$W^{(k)\alpha}$}}
\newcommand{\Fmv} {\mbox{$\mathrm{F}^{\mu\nu}$}} 
\newcommand{\FMV} {\mbox{$\mathrm{F}_{\mu\nu}$}} 
\newcommand{\Fma} {\mbox{$\mathrm{F}^{\mu\alpha}$}} 
\newcommand{\FMB} {\mbox{$\mathrm{F}_{\mu\beta}$}} 

\newcommand{\WM} {\mbox{     $\vec{{W}}_{\mu}$  }}
\newcommand{\Wa} {\mbox{     $\vec{{W}}^\alpha$ }}
\newcommand{\WA} {\mbox{     $\vec{{W}}_\alpha$ }}

\newcommand{\Wb} {\mbox{     $\vec{{W}}^\beta$  }}

\newcommand{\WV} {\mbox{$W_{\nu}$}}

\newcommand{\WWg}{\mbox{\WW$\gamma$}}
\newcommand{\sigWWg}{\mbox{$\hat{\sigma}_{\mathrm{WW}\gamma}$}}
\newcommand{\effWWg}{\mbox{$\varepsilon_{\mathrm{WW}\gamma}$}}
\newcommand{\PWWg}{\mbox{$P_{\mathrm{WW}\gamma}$}}
\newcommand{\sigback}{\mbox{$\sigma_{\mathrm{BGD}}$}}

\newcommand{\cres}{\mbox{$c_{\mathrm{res}}$}}

\newcommand{\ZZ}{\mbox{\Zzero\Zzero}}

\newcommand{\qq}{\mbox{$\mathrm{q\overline{q}}$}}

\newcommand{\qqg}{\mbox{$\mathrm{q\overline{q}}\gamma$}}

\newcommand{\lnu}{\mbox{$\ell\overline{\nu}_{\ell}$}}

\newcommand{\enu}{\mbox{$\mathrm{e\overline{\nu}_{e}}$}}
\newcommand{\mnu}{\mbox{$\mu\overline{\nu}_{\mu}$}}
\newcommand{\tnu}{\mbox{$\tau\overline{\nu}_{\tau}$}}
\newcommand{\WWqqqq}{\mbox{\WW$\rightarrow$\qq\qq}}
\newcommand{\WWgqqqq}{\mbox{\WW$(\gamma)\rightarrow$\qq\qq$\gamma$}}

\newcommand{\qqqq}{\mbox{\qq\qq}}

\newcommand{\qqlv}{\mbox{\qq\lnu}}

\newcommand{\lnln}{\mbox{\lnu\lnu}}
\newcommand{\WWqqln}{\mbox{\WW$\rightarrow$\qq\lnu}}
\newcommand{\WWgqqln}{\mbox{\WW$(\gamma)\rightarrow$\qq\lnu$\gamma$}}
\newcommand{\WWqqen}{\mbox{\WW$\rightarrow$\qq\enu}}
\newcommand{\Wev}{\mbox{$\mathrm{W}$\enu}}
\newcommand{\WWqqmn}{\mbox{\WW$\rightarrow$\qq\mnu}}
\newcommand{\WWqqtn}{\mbox{\WW$\rightarrow$\qq\tnu}}

\newcommand{\Zqq}{\mbox{\Zzero$\rightarrow$\qq}}
\newcommand{\Mzs}{\mbox{$M^2_{\mathrm{Z}^0}$}}
\newcommand{\Mz}{\mbox{$M_{\mathrm{Z}^0}$}}
\newcommand{\Mw}{\mbox{$M_{\mathrm{W}}$}}

\newcommand{\Mws}{\mbox{$M^2_{\mathrm{W}}$}}
\newcommand{\Mfafb}{\mbox{$M_{f_1\overline{f}_2}$}}
\newcommand{\Mfcfd}{\mbox{$M_{f_3\overline{f}_4}$}}
\newcommand{\fafb}{\mbox{$f_1\overline{f}_2$}}
\newcommand{\fcfd}{\mbox{$f_3\overline{f}_4$}}
\newcommand{\Gw}{\mbox{$\Gamma_{\mathrm{W}}$}}

\newcommand{\Egam}{\mbox{$E_\gamma$}}

\newcommand{\Opal}{\mbox{OPAL}}

\newcommand{\cosg}{\mbox{$\cos\theta_\gamma$}}
\newcommand{\cosgl}{\mbox{$\cos\theta_{\gamma \ell}$}}
\newcommand{\cosgf}{\mbox{$\cos\theta_{\gamma f}$}}
\newcommand{\cosgj}{\mbox{$\cos\theta_{\gamma-\mathrm{JET}}$}}

\newcommand{\Misr}{\mbox{$M_{\mathrm{ISR}}$}}

\newcommand{\Mfsr}{\mbox{$M_{\mathrm{FSR}}$}}

\newcommand{\roots}{\mbox{$\sqrt{s}$}}
\newcommand{\rootsprime}{\mbox{$\sqrt{s'}$}}

\newcommand{\KoralW}{\mbox{KORALW}}

\newcommand{\PHOTOS}{\mbox{PHOTOS}}
\newcommand{\KORALW}{\mbox{KORALW}}

\newcommand{\KKFF}{\mbox{KK2F}}

\newcommand{\Excalibur}{\mbox{EXCALIBUR}}

\newcommand{\grcff}{\mbox{GRC4F}}
\newcommand{\Pythia}{\mbox{PYTHIA}}

\newcommand{\Racoon}{\mbox{RACOONWW}}

\newcommand{\WWF}{\mbox{WWF}}

\newcommand{\Herwig}{\mbox{HERWIG}}
\newcommand{\Jetset}{\mbox{JETSET}}

\newcommand{\eewwgammamc}{\mbox{EEWWG}}
\newcommand{\eewwg}{\mbox{EEWWG}}
\newcommand{\EEWWG}{\mbox{EEWWG}}

\def\etal{\mbox{{\it et al.}}}
\documentstyle[a4p,12pt,epsfig,amssymb]{article}
\begin{document}

\begin{titlepage}

\begin{center}
{\large EUROPEAN LABORATORY FOR PARTICLE PHYSICS}
\end{center}
\bigskip
\begin{flushright}
CERN-EP/99-130 \\ 22 September 1999
\end{flushright}
\bigskip\bigskip
\begin{center}
 \huge{\bf \boldmath Measurement of the  \\} 
 \huge{\bf \boldmath \WWg\ Cross-section and\\ }
 \huge{\bf   First Direct Limits on  Anomalous \\} 
 \huge{\bf \boldmath Electroweak Quartic Gauge Couplings } 
\end{center}
\bigskip\bigskip\bigskip
\begin{center}
{ \LARGE The OPAL Collaboration}
\vspace{0.25cm}
\end{center}
\bigskip\bigskip\bigskip

\begin{center}{\large Abstract}\end{center}
A study of \WW\ events accompanied by hard photon radiation 
produced in \epem\ collisions at LEP is presented. Events consistent
with two on-shell W-bosons and an isolated photon are selected from
183~pb$^{-1}$ of data recorded at $\roots=189$~GeV. 
From these data, 17 \WWg\ candidates are selected with 
photon energy greater than 10~GeV, consistent with the Standard Model
expectation. These events are used to measure the 
$\epem\rightarrow\WWg$ cross-section within a set of geometric and 
kinematic cuts,
$\sigWWg = 
136 \pm 37 \pm 8$~fb, where the first error is statistical and the
second systematic. The photon energy spectrum is used to set the first
direct, albeit weak, limits on possible anomalous contributions to the
$\WW\gamma\gamma$ and $\WW\gamma\Zzero$ vertices:
\begin{eqnarray*}
   -0.070~\mathrm{GeV}^{-2} < & a_0/ \Lambda^2 & < 0.070~\mathrm{GeV}^{-2}, \\ 
    -0.13~\mathrm{GeV}^{-2} < & a_c/ \Lambda^2 & < 0.19~\mathrm{GeV}^{-2}, \\
    -0.61~\mathrm{GeV}^{-2} < & a_n/ \Lambda^2 & < 0.57~\mathrm{GeV}^{-2},  
\end{eqnarray*}
where $\Lambda$ represents the energy scale for new physics. 

\bigskip\bigskip\bigskip
%%\begin{center}
%% {\huge\bf FINAL DRAFT}
%%\end{center}
%%\begin{center}
%% {\large\bf Comments to Mark.Thomson@cern.ch before 0900 22/09/99} 
%%\end{center}
\bigskip\bigskip\bigskip

\begin{center}
  {\large Submitted to Physics Letters B}
\end{center}
\end{titlepage} 

\begin{center}{\Large        The OPAL Collaboration
}\end{center}\bigskip
\begin{center}{
%begin authorlist PLEASE DO NOT DELETE THIS COMMENT
G.\thinspace Abbiendi$^{  2}$,
K.\thinspace Ackerstaff$^{  8}$,
P.F.\thinspace Akesson$^{  3}$,
G.\thinspace Alexander$^{ 23}$,
J.\thinspace Allison$^{ 16}$,
K.J.\thinspace Anderson$^{  9}$,
S.\thinspace Arcelli$^{ 17}$,
S.\thinspace Asai$^{ 24}$,
S.F.\thinspace Ashby$^{  1}$,
D.\thinspace Axen$^{ 29}$
% ...
%end authorlist PLEASE DO NOT DELETE THIS COMMENT
}\end{center}\bigskip
\bigskip
%begin institutes
$^{  1}$School of Physics and Astronomy, University of Birmingham,
Birmingham B15 2TT, UK
\newline
$^{  2}$Dipartimento di Fisica dell' Universit\`a di Bologna and INFN,
I-40126 Bologna, Italy
\newline
$^{  3}$Physikalisches Institut, Universit\"at Bonn,
D-53115 Bonn, Germany
\newline
% ...
$^{  8}$CERN, European Organisation for Particle Physics,
CH-1211 Geneva 23, Switzerland
\newline
$^{  9}$Enrico Fermi Institute and Department of Physics,
University of Chicago, Chicago IL 60637, USA
\newline
% ...
$^{ 16}$Department of Physics, Schuster Laboratory, The University,
Manchester M13 9PL, UK
\newline
$^{ 17}$Department of Physics, University of Maryland,
College Park, MD 20742, USA
\newline
% ...
$^{ 23}$Department of Physics and Astronomy, Tel Aviv University,
Tel Aviv 69978, Israel
\newline
$^{ 24}$International Centre for Elementary Particle Physics and
Department of Physics, University of Tokyo, Tokyo 113-0033, and
Kobe University, Kobe 657-8501, Japan
\newline
% ...
$^{ 29}$University of British Columbia, Department of Physics,
Vancouver BC V6T 1Z1, Canada
\newline
% ...
%end notes
 
\section{Introduction}

\begin{figure}[h]
 \begin{center}
 \epsffile{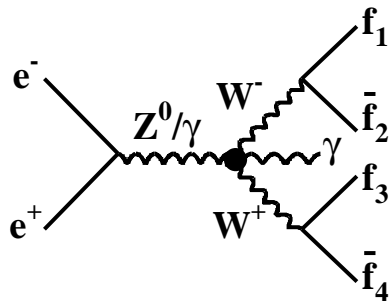}
 \caption{Standard Model $\WW\gamma\gamma$ and $\WW\Zzero\gamma$ 
          quartic gauge couplings}
 \label{fig:qgcdiag}
 \end{center}
\end{figure}

The \WW\ cross-section has been precisely measured at LEP over a range of
centre-of-mass energies~\cite{bib:xs183,bib:lepxs}, 
and is well described by the 
Standard Model (SM) expectation~\cite{bib:gentle}. 
In this paper these measurements are extended 
to include also \WW\ events with energetic photons in order
to probe the modelling of electro-magnetic radiation in the \WW\ pair 
creation process.
No previous measurements exist of three vector boson production when at 
least two bosons are massive. In addition, the \WWg\ final state 
may be sensitive to anomalous contributions to the SM 
$\WW\gamma\gamma$ and $\WW\Zzero\gamma$ quartic gauge couplings, shown in
Figure \ref{fig:qgcdiag}.

The non-Abelian nature of the electroweak sector of the Standard Model  
results in vector boson self-interactions. In addition to the
triple gauge boson couplings (TGCs), $\WW\gamma$ and $\WW\Zzero$, 
the Standard Model predicts the existence of four quartic gauge couplings, 
$\WW\WW$, $\WW\ZZ$, $\WW\Zzero\gamma$ and $\WW\gamma\gamma$. 
These couplings are not expected to play a significant role at LEP
energies, but will be important at a future TeV linear 
collider~\cite{bib:Boos} and at the
LHC~\cite{bib:LHC}. There has been neither direct 
experimental confirmation of the existence of quartic couplings, nor
any  direct limits on possible anomalous quartic couplings. 
However, indirect limits on anomalous quartic couplings
can be derived from 
the precise LEP/SLD $\Zzero$ data~\cite{bib:Eboli,bib:Brunstein}. 
 
Quartic gauge boson couplings can be probed in final states with three vector
bosons. At LEP centre-of-mass energies, final states involving three massive
gauge bosons are kinematically out of reach. However, it is possible to 
study the processes $\epem\rightarrow\WW\gamma$ and
$\epem\rightarrow\Zzero\gamma\gamma$. In the Standard Model,
the contribution of the quartic coupling to $\epem\rightarrow\WW\gamma$ is
expected to be too small to measure and that to 
$\epem\rightarrow\Zzero\gamma\gamma$ is zero. Nevertheless, it is possible
to set the first direct limits on possible anomalous contributions to
the quartic gauge boson couplings.

This paper describes a study of the process 
$\epem\rightarrow\WW\gamma$ using 183~pb$^{-1}$ of data recorded at a
centre-of-mass energy of 189~GeV with the \Opal\ detector at LEP. 
It begins with an introduction to the phenomenology of 
quartic gauge boson couplings at LEP, followed by a description
of the selection of $\WW\gamma$ events and of the measurement of the 
$\WW\gamma$ cross-section for photon energies $\Egam > 10$~GeV.
The results are then interpreted as direct limits on possible anomalous
$\WW\gamma\gamma$ and $\WW\Zzero\gamma$ quartic gauge couplings and
compared with the indirect limits.

\section{Anomalous Quartic Gauge Couplings}

In the Standard Model the form and strength of the vector boson 
self-interactions are fixed by SU(2)$\times$U(1) gauge invariance.
As is the case for triple gauge boson couplings~\cite{bib:Hagiwara}, 
in extensions to the Standard Model, anomalous quartic couplings
can be parametrised by additional terms in the 
Lagrangian~\cite{bib:Berlanger,bib:Stirling1,bib:Stirling2}.
These are required to conserve 
custodial SU$(2)_c$ symmetry in order to avoid deviations of the 
$\rho$ parameter\footnote{$\rho = \Mws/(\Mzs\cos^2\theta_W)$, where \Mw\ 
and \Mz\ are the masses of the W$^\pm$ and \Zzero\ bosons and 
$\theta_W$ is the weak mixing angle.} 
from the experimentally well established value close to 1. 
Only operators which do not introduce anomalous triple gauge 
couplings are considered. For example, the anomalous quadrupole moment 
operator generates both \WW$\gamma$ and \WW$\gamma\gamma$ couplings. 
Therefore, it is not considered as a source of genuine anomalous quartic 
couplings since its strength, $\lambda_\gamma$, is already tightly 
constrained from the study 
of TGCs at LEP~\cite{bib:xs183,bib:leptgc} and at 
the Tevatron~\cite{bib:tevatgc}. The lowest dimension operators 
which generate genuine anomalous quartic couplings involving photons are 
of dimension six. Three such possibilities are considered here, 
${\cal{L}}_6^0$, 
${\cal{L}}_6^c$~\cite{bib:Berlanger} and 
${\cal{L}}_6^n$~\cite{bib:Eboli,bib:Stirling2}:
\begin{eqnarray*}
     {\cal{L}}_6^0 & = & 
             -{{e^2}\over{16\Lambda^2}}a_0 \Fmv \FMV \Wa.\WA, \\
     {\cal{L}}_6^c & = & 
             -{{e^2}\over{16\Lambda^2}}a_c \Fma \FMB \Wb.\WA, \\
     {\cal{L}}_6^n & = &       
              i{{e^2}\over{16\Lambda^2}}a_n \epsilon_{ijk} \Wma^{(i)} 
               \WV^{(j)}\Wka\Fmv, 
\end{eqnarray*}
with 
\begin{eqnarray*}
     \WM & = & \left(  \begin{array}{c}
                         \frac{1}{\sqrt{2}}( W_\mu^+ + W_\mu^-) \\ 
                         \frac{i}{\sqrt{2}}( W_\mu^+ - W_\mu^-) \\
                         Z_\mu / \cos\theta_{W} 
                       \end{array}
               \right),
\end{eqnarray*}
where $\Fmv$ and $\Wmv$ are the field strength tensors of the
photon and $W$ fields respectively. 
Both ${\cal{L}}_6^0$ and ${\cal{L}}_6^c$, which conserve 
$C$ and $P$ (separately), generate 
anomalous $\WW\gamma\gamma$ and $\ZZ\gamma\gamma$ couplings. 
The $CP$ violating term
${\cal{L}}_6^n$ results in an anomalous $\WW\Zzero\gamma$ 
coupling\footnote{In reference \cite{bib:lep2boudjema} it is pointed
out that the form for ${\cal{L}}_6^n$ used in 
\cite{bib:Stirling2} is missing a factor of $i$ which is crucial
for hermiticity. The missing factor of $i$ was a typographical error
in the paper. However, the implementation in the EEWWG 
program is correct\cite{bib:oops}.}.  
In each case, the strength of the
coupling is proportional to $a_i/ \Lambda^2$, where $\Lambda$ 
represents a scale for new physics. 
A more general description of the operators leading to 
anomalous quartic couplings accessible at LEP can be found 
in the recent paper of B\'elanger \etal\cite{bib:lep2boudjema}. 

\begin{figure}[htbp]
 \begin{center}
  \epsfxsize=\textwidth
  \epsffile{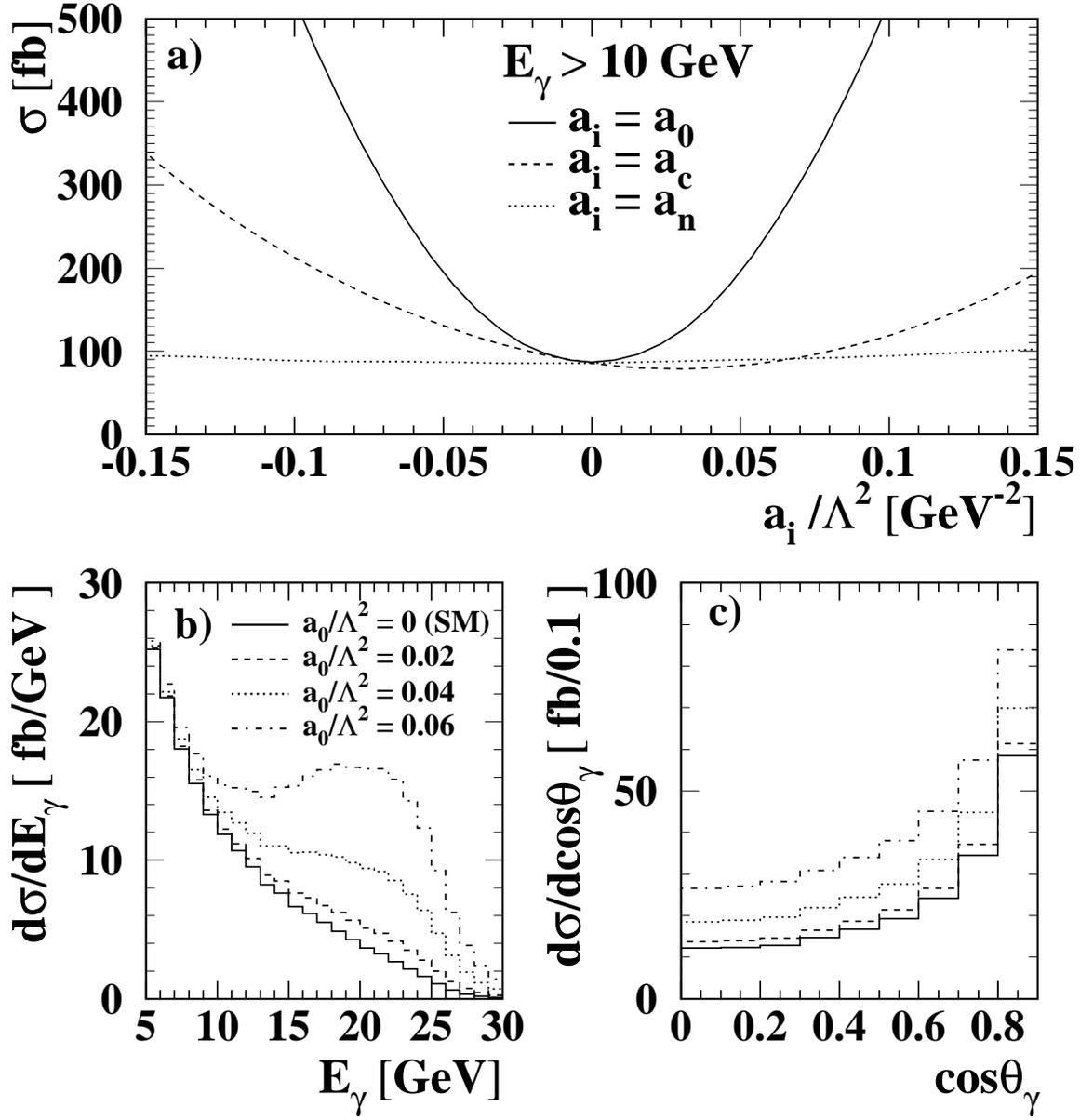}
  \caption{The effect of anomalous quartic couplings in 
   $\epem$ collisions at $\roots=$189~GeV.
   a) Expected $\WW\gamma$ cross-section for different 
   anomalous quartic couplings, the photon was required to have
   energy, $\Egam>10$~GeV and be within the polar acceptance, $|\cosg|<0.9$.
   The kinematic cuts used to define the cross-section, see text, were
   also applied. Plots b) and c) show the effect of an anomalous quartic gauge 
   coupling, $a_0$, on the energy and angular distribution of photons in 
   $\WW\gamma$ events, where the fiducial cuts, $\Egam>5$~GeV and 
   $|\cosg|<0.9$ were imposed.
   In all three figures, the contribution from final state
   radiation diagrams is not included.}
   \label{fig:astudy}
 \end{center}
\end{figure}

Figure~\ref{fig:astudy}a shows the predicted cross-section for 
$\WW\gamma$ as a function of different values of 
$a_0/\Lambda^2$, $a_c/\Lambda^2$ and $a_n/\Lambda^2$, using the
calculation of Stirling and Werthenbach~\cite{bib:Stirling2}. 
The forms of ${\cal{L}}_6^0$, ${\cal{L}}_6^c$ and ${\cal{L}}_6^n$ 
dictate that contributions to the $\WWg$ matrix element from
the anomalous quartic couplings scale linearly with 
the energy of the photon~\cite{bib:Stirling2}. 
The cross-section calculation of reference~\cite{bib:Stirling2} 
has been used to investigate further the effects of 
anomalous couplings on the energy spectrum of photons in $\WW\gamma$ events 
and on the angular distribution of the photons.
For $\roots=189$~GeV the effects of a large anomalous coupling 
({\em e.g.} $a_0$) are shown in 
Figure~\ref{fig:astudy}b and \ref{fig:astudy}c. 
The signal of anomalous quartic gauge boson couplings at LEP would
be an excess of \WWg\ events with photon energy greater than
10~GeV. 

%%%%%%%%%%%%%%%%%%%%%%%%%%%%%%%%%%%%%%%%%%%%%%%%%%%%%%%%%%%%%%%%%%%%%%%%%%%%%

\section{\boldmath The $\WW\gamma$ Final State }
\label{sec:WWG}

There are 23 lowest order diagrams which contribute to the process 
$\epem\rightarrow\WW(\gamma)\rightarrow\qqlv\gamma$. 
There are an additional 3 diagrams for the process 
$\epem\rightarrow\WW(\gamma)\rightarrow\qqqq\gamma$. 
These diagrams correspond to initial state
radiation (ISR), final state radiation (FSR), radiation from the W-boson
(WR) and the Standard Model quartic gauge coupling (QGC) diagram.
Different Monte Carlo generators  and calculations employ different 
subsets/implementations of these diagrams.
In this study, two Monte Carlo generators, \WWF~\cite{bib:WWF} 
and \KoralW~\cite{bib:KORALW} are used to 
simulate the Standard Model expectation for the $\WW\gamma$ final state.
The \WWF\ Monte Carlo generator produces
$\WW\gamma$ events above some photon energy cut and 
within a specified acceptance in polar angle. This program implements a
full matrix element calculation for the Standard 
Model ISR, FSR, WR and QGC diagrams~\cite{bib:Aeppli}. 
However, this treatment may not be optimal for radiation from quarks 
where it is also necessary to consider QCD corrections to the
radiation process.
Higher order corrections are implemented using a structure function 
approach to collinear ISR.
The \KoralW\ generator, used to simulate $\WW(\gamma)$ events,  
does not include contributions
from radiation from the intermediate W-bosons  or
from the Standard Model quartic coupling diagram. 
FSR  from leptons is implemented using the 
\PHOTOS~\cite{bib:Photos} program and FSR from quarks is modelled with 
\Jetset~\cite{bib:Jetset}. However, since the Standard Model $\WWg$ 
cross-section is dominated by ISR, reasonable predictions are obtained.
The Monte Carlo samples generated with both \WWF\ and \KoralW\ 
are passed through
the full \Opal\ detector simulation~\cite{bib:GOPAL}.
Neither \WWF\ nor \KoralW\ currently allows for  
anomalous quartic gauge couplings.
The calculation of Stirling and Werthenbach~\cite{bib:Stirling2} 
allows for the assessment of the impact of anomalous quartic couplings
and is implemented in the \eewwg\ program.
This calculation includes the ISR diagrams, the WR diagrams, the
SM quartic diagram and can accommodate anomalous quartic couplings. 
It does not include FSR and therefore ignores any possible interference
effects between the FSR diagrams and the other diagrams. 

\subsection{\boldmath $\WW\gamma$ Signal Definition}
\label{sec:sigdef}

The process  $\epem\rightarrow\WW\gamma$ results in a four-fermion plus 
photon final state, $\fafb\fcfd\gamma$, where the fermion flavours are
appropriate for W-decay. Two distinct regions in phase space are considered. 
In the first, the $\fafb$ and $\fcfd$ systems are 
produced with invariant masses close to \Mw. In the second region, 
which is dominated by final state radiation, either 
the $\fafb\gamma$ and $\fcfd$ 
combinations or the $\fafb$ and $\fcfd\gamma$ combinations give 
invariant masses close to the W-boson mass. The second region has
little sensitivity to anomalous quartic couplings. For photon energies
$\Egam\gg\Gw$, where $\Gw$ is the W-boson width, it is possible to 
separate the phase space of events from FSR from those arising from 
the ISR or QGC diagrams using the invariant masses of the $\fafb$ and 
$\fcfd$ systems.
 
Both the selection procedure and the signal definition are designed to 
suppress contributions from FSR which is 
effectively a background process. The cuts
used also reject photons from the decays of 
$\pi^0$ and $\eta$ mesons associated with the hadronic jets. 
For this paper, the 
$\WW\gamma\rightarrow\fafb\fcfd\gamma$ cross-section, denoted by
$\sigWWg$, is defined by: 
\begin{itemize}
  \item $|\cosg| < 0.9$, where $\cosg$ is the cosine of the polar 
                         angle\footnote{The \Opal\ right-handed
                         coordinate system is defined such that the
                         origin is at the centre of the detector and the
                         $z$ axis points along the direction of the $e^-$
                         beam, $\theta$ is the polar angle with respect to 
                         the $z$ axis.} of the photon.
  \item $\cosgf  < 0.9$, where $\cosgf$ is the cosine of the
                         minimum angle between the
                         photon and any of the charged fermions in the
                         four-fermion final state.
  \item minimum($\Mfafb,\Mfcfd) > 73$~GeV, where \Mfafb\ and \Mfcfd\ are
                              the invariant masses of the fermions from 
                              the W$^-$ and from the W$^+$.
  \item $\Egam   > 10$~GeV, where \Egam\ is the photon energy.
\end{itemize}
In the above definition, all cuts are made on generator level quantities.
Generator level refers to the true four-momenta of particles in the 
$\fafb\fcfd\gamma$ final state.
The first two cuts suppress ISR and FSR respectively and the 
invariant mass cuts further suppress final state radiation. 
%In a sample of \KoralW\ Monte Carlo \WW\ events
%(where ISR and FSR can be unambiguously separated from the Monte Carlo
%generator information), the mass cuts  retain 87\% of events with 
%ISR photons with $\Egam>10$~GeV, $|\cosg| < 0.9$, and $\cosgf< 0.9$. 
%The contamination from FSR in this region is 2.5\%. 
The cross-section within the above kinematic cuts,
\sigWWg, is dominated
by doubly-resonant \WW\ production.
At $\roots=189$~GeV, the effects of other four-fermion diagrams
resulting in final states which can interfere with $\WW\gamma$
have been evaluated with the \Racoon\ program~\cite{bib:Denner}.
For the signal acceptance, the cross-section
for the $\epem\rightarrow\WWg$ diagrams alone is 
$109.33\pm0.02$~fb and when all 
(interfering) four-fermion diagrams are included this becomes 
$109.84\pm0.03$~fb~\cite{bib:Doreen}, where in both cases
the errors are statistical. Currently the predictions of 
\Racoon\ do not include higher-order corrections from 
collinear ISR. 

The Standard Model expectations for the cross-section
for the $\WW\gamma$ final state within this 
acceptance for $\Egam>10$~GeV have been determined 
using the \WWF\ and \KoralW\ generators and the
\eewwgammamc\ program. The results are summarised in 
Table~\ref{tab:smxsec}. Also given in Table~\ref{tab:smxsec} are the 
cross-sections from \eewwg, \WWF\ and \Racoon\ where the effect of collinear 
ISR has been neglected.    

\section{The OPAL Detector, Data and Monte Carlo Samples }

The \Opal\ detector includes a 3.7 m diameter tracking volume within
a 0.435~T axial
magnetic field. The tracking detectors consist of a silicon
micro-vertex detector, a high precision gas vertex detector and a large
volume gas jet chamber. Lying outside the solenoid, 
the electro-magnetic calorimeter consisting of $11\,704$ lead glass blocks
has  full acceptance in the range $|\cos\theta|<0.98$ and a relative
energy resolution of approximately $6\%$ for 10~GeV photons.
The magnet return yoke is instrumented with streamer tubes which
serve as the hadronic calorimeter. Muon chambers outside the 
hadronic calorimeter provide muon identification in the range
$|\cos\theta|<0.98$.  A detailed description of the \Opal\ detector can be
found in \cite{bib:detector}.

The integrated luminosity of the data sample recorded during the 1998
LEP operation, evaluated using small angle
Bhabha scattering events observed in the silicon tungsten forward 
calorimeter~\cite{bib:lumi}, is
$183.05\pm0.16(\mathrm{stat.})\pm0.37(\mathrm{sys.})~\mathrm{pb}^{-1}$.
The luminosity-weighted mean centre-of-mass energy 
for the data sample is $\roots=188.63\pm0.04$~GeV. 

Both the \WWF\ and \KoralW\ Monte Carlo samples are used to simulate
\WWg\ events. Selection efficiencies are estimated using \WWF. 
Differences between efficiencies calculated using \WWF\ and \KoralW\
are assigned as systematic uncertainties related to the theoretical
modelling of \WWg\ production, and in particular, the modelling of
FSR. A number of Monte Carlo samples, all including a full simulation of the
\Opal\ detector, are used to simulate background processes to $\WW(\gamma)$.
The background process \Zqq\ is simulated using 
\Pythia~\cite{bib:pythia}, with \Herwig~\cite{bib:Herwig} and 
\KKFF~\cite{bib:kk} 
used as alternatives to assess systematic uncertainties. The
four-fermion backgrounds from $(\Zgamma)(\Zgamma)$ and \Wev\ are
simulated with \Pythia, \grcff~\cite{bib:GRC4F} and 
\Excalibur~\cite{bib:excalibur}.

\section{\boldmath $\WW\gamma$ Event Selection}

The selection of $\WW\gamma$ events proceeds in three stages:
selection of $\WW$ events, photon identification and acceptance cuts,
and kinematic requirements. Initially the energy requirement on
identified photons is $\Egam>5$~GeV. However, for the determination of
the $\WWg$ cross-section and the anomalous coupling analysis only
events with photon energies $\Egam>10$~GeV are used. The photon energy
range 5-10~GeV has little sensitivity to anomalous couplings and 
is subject to a larger background contribution and consequently larger 
systematic uncertainties.
 
For this analysis, 
$\WW(\gamma)\rightarrow\lnln\gamma$ events are not used because
it is not possible to reject adequately FSR photons using the kinematics
of the event. However, in the determination of $\sigWWg$ 
the value quoted has been corrected to include
all final states.

\subsection{\boldmath \WW\ Selection}

The \WWqqln\ and \WWqqqq\ selections of reference \cite{bib:xs183}
are used as the basis of the \WWg\ selection at 189~GeV. 
For the selection of $\WWgqqln$ events, the standard $\WWqqln$ 
selection is applied, but with one 
additional requirement. 
In a significant fraction of events selected as 
$\WW\rightarrow\qq(\tau\rightarrow  h^\pm n\pi^0\nu)\nu$ and
$\WW\rightarrow\qq(\tau\rightarrow 3h^\pm n\pi^0\nu)\nu$,
the track(s) and clusters identified as the tau decay products
are in reality associated with one of the jets.
In order to reduce this contamination, the tau decay candidate
track (highest momentum track in the case of 3-prong decays) is required 
to have momentum
greater than 3 GeV. For photons with $|\cosg|<0.9$ 
the selection efficiency for \WWgqqln\ events 
is approximately constant at 87\% in the photon energy
range 10-30~GeV, and is almost independent of the 
polar angle of the photon within the accepted region.

For the selection of $\WWgqqqq$ events, a modified version of
the $\WWqqqq$ selection of reference \cite{bib:xs183} is used. 
In this selection events are forced into four jets using the Durham
$k_{T}$ algorithm~\cite{bib:Durham}. In 
approximately $10\%$ of Monte Carlo events with 
a photon with $\Egam>10$~GeV, the photon alone forms one of the
four jets. This introduces an additional inefficiency, due to the
requirement in the preselection that there should be at least
one charged track associated with each jet. 
For this reason, events failing the standard
$\WWqqqq$ selection are forced into four jets after excluding the highest
energy isolated electro-magnetic calorimeter cluster and the selection
re-applied.
The  selection efficiency for $\WWqqqq\gamma$ events 
with photon energies in the range 10-30~GeV is 90\% and 
is approximately independent of the photon energy and photon polar angle.
 
\subsection{Photon Identification and Acceptance} 

Photon identification is similar to that described in \cite{bib:OPAL_HGG}.
Photon candidates are identified as one of three types:
\begin{itemize}
 \item Unassociated electro-magnetic calorimeter clusters 
       defined by the requirement that no 
       charged track lie within the angular resolution of the cluster 
       when extrapolated to the calorimeter. The lateral spread of the 
       cluster was required to satisfy the criteria described in 
       reference \cite{bib:OPAL_HGG}. 
 \item Two-track photon conversions which 
       are selected using an artificial 
       neural network.
 \item Conversions where only a single track is reconstructed, identified 
       as an electro-magnetic calorimeter 
       cluster associated with a charged track which is consistent 
       with pointing to the primary vertex. The track is
       required to have no associated hits 
       in either layer of the silicon micro-vertex detector or in the 
       first 6 layers of the central vertex chamber.
       Up to one additional charged track passing the
       same criteria is allowed to point to the cluster.  
\end{itemize}
For both types of conversions, the photon energy is defined by the 
sum of cluster energies pointed to by the track(s).

Photon candidates identified using the above criteria are 
required to satisfy isolation requirements. 
The energy of additional tracks and clusters in a $20^\circ$ half-angle cone 
defined by the photon direction has to be less than 2 GeV. In addition,
the energy deposited in the hadron calorimeter in a $20^\circ$ half-angle 
cone around the photon candidate is required to be less than 5~GeV.

The identified photon is required to lie within the polar acceptance,
\begin{itemize}
 \item $|\cosg| < 0.9$.
\end{itemize}
The photon is also required to be isolated from the charged
fermions in the final state.
Cuts are applied on the cosine of the angle between the 
photon and closest jet, $\cosgj$, and on the cosine of the 
angle between the photon and lepton, $\cosgl$:
\begin{itemize}
 \item $\cosgj < 0.9$,
 \item $\cosgl < 0.9$ for $\WWqqen\gamma$ and $\WWqqmn\gamma$,
 \item $\cosgl < 0.7$ for $\WWqqtn\gamma$.
\end{itemize}
For photons within the generator level acceptance
$\Egam>10$~GeV, $|\cosg|<0.9$ and
$\cosgf<0.9$, the photon identification
efficiency is about 83\% for selected $\WWgqqln$ and $\WWgqqqq$ events.

\subsection{Kinematic Requirements}

Kinematic cuts which reject events with FSR photons or with 
photons associated with the hadronic jets, {\em i.e.}
from $\pi^0$ and $\eta$ decays, are performed by comparing
the results of three kinematic fits.
In each case, the constraints of energy and momentum conservation are
imposed. The finite W width is neglected and the two 
reconstructed masses of the W boson candidates are required to be 
equal~\cite{bib:mass172}. 
The fits employed correspond to the following hypotheses: 
\begin{itemize}
 \item[a)] FSR(quark), assuming a two-body \WW\ final state, where the identified photon 
       is associated with one of the jets.
 \item[b)] FSR(lepton),  
       only used for \WWgqqln\ events, assuming a two-body \WW\ final state, 
       where the photon is associated with the charged lepton.
 \item[c)] ISR/QGC, assuming a three body final state 
       consisting of two W bosons and the photon. 
\end{itemize}  
An event is considered consistent with one of the above
fit hypotheses if the fit converges with a fit probability of greater than
0.1\% and if the reconstructed W boson mass is greater than 74~GeV.
Only \WWg\ candidate 
events consistent with being two approximately on-shell W-bosons (fit $c$),
$\Mfafb\sim\Mfcfd\sim\Mw$, are retained for the measurement
of the \WWg\ cross-section and quartic gauge coupling study. This 
procedure suppresses events with final state radiation and events
where the photon is from the decay of mesons.
By reducing contamination from FSR, the sensitivity to
anomalous quartic couplings is improved.
In addition, the systematic uncertainties from 
photons associated with jets (FSR and $\pi^0$/$\eta$ decays)
are significantly reduced. 
%%The separation between ISR and FSR 
%%increases with increasing photon energy.
 
\begin{figure}[htbp]
 \epsfxsize=\textwidth
 \epsffile{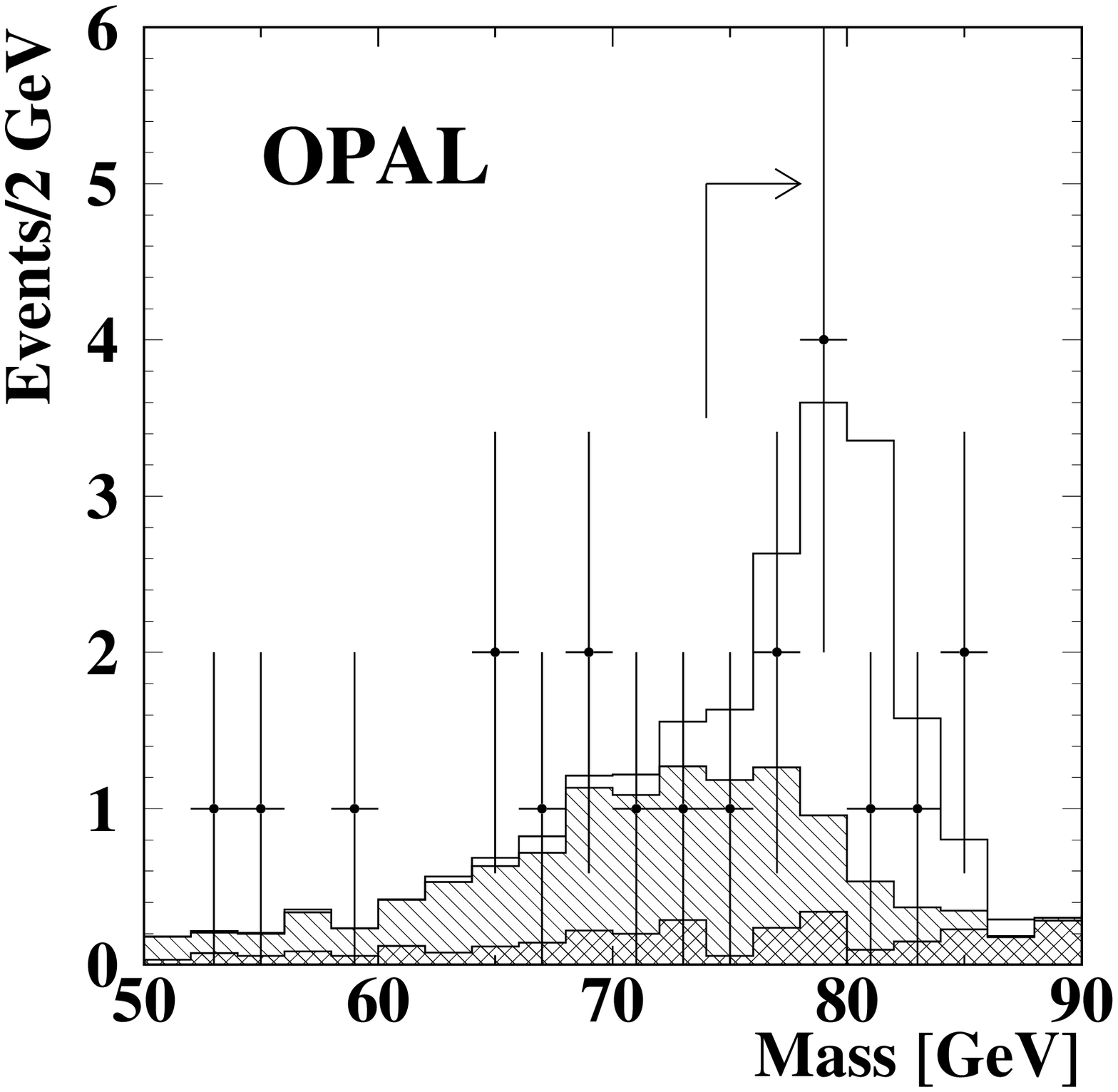}
 \caption{The reconstructed W-boson mass distribution for candidate 
          \WWgqqln\ events using the ISR/QGC hypothesis. The points show
          the data. The histogram shows the Monte Carlo expectation where
          \KoralW\ has been used to simulate the $\WW(\gamma)$ final state
          and \Pythia\ is used to simulate non-\WW\ background. The hatched
          histogram shows the contribution from \WW\ events with final state
          radiation or where the photon candidate originates from the
          decay of a meson. The double-hatched histogram shows the
          contribution from non-\WW\ background. The cut 
          is indicated by the arrow.}
 \label{fig:isrcut}
\end{figure}

Selected \WWgqqln\ events are required to be consistent with the ISR/QGC
hypothesis  using the above criteria. The effect of the invariant mass cut
is indicated in Figure \ref{fig:isrcut}.
In order further to reduce contributions from FSR, if the event is also 
consistent with the hypothesis of FSR from the lepton, it is required that
the reconstructed mass from the ISR/QGC fit be closer to \Mw\ than the
mass obtained from the FSR(lepton) fit.

In fully hadronic events there are three possible jet-pairing 
combinations. Only events where one of the three combinations is
kinematically consistent with the ISR/QGC hypothesis are retained. 
If more than one combination satisfies this requirement, the fit yielding
the highest probability is used. Events which are also consistent with
the FSR(quark) hypothesis are rejected if the
FSR fit has higher probability than the ISR/QGC fit
and if $\Mfsr < 86 $~GeV and $\Misr < 78$~GeV, where \Mfsr\ and \Misr\ are
the respective reconstructed masses. 

The application of the above kinematic requirements 
retains approximately 80\% of \WWg\ events within the signal
definition of Section \ref{sec:sigdef} whilst rejecting 
85\%-98\% (increasing with the photon energy) of events with photons 
from FSR or from the decays of mesons.

\section{\boldmath \WWg\ cross-section}

\begin{figure}[htbp]
 \epsfxsize=\textwidth
 \epsffile{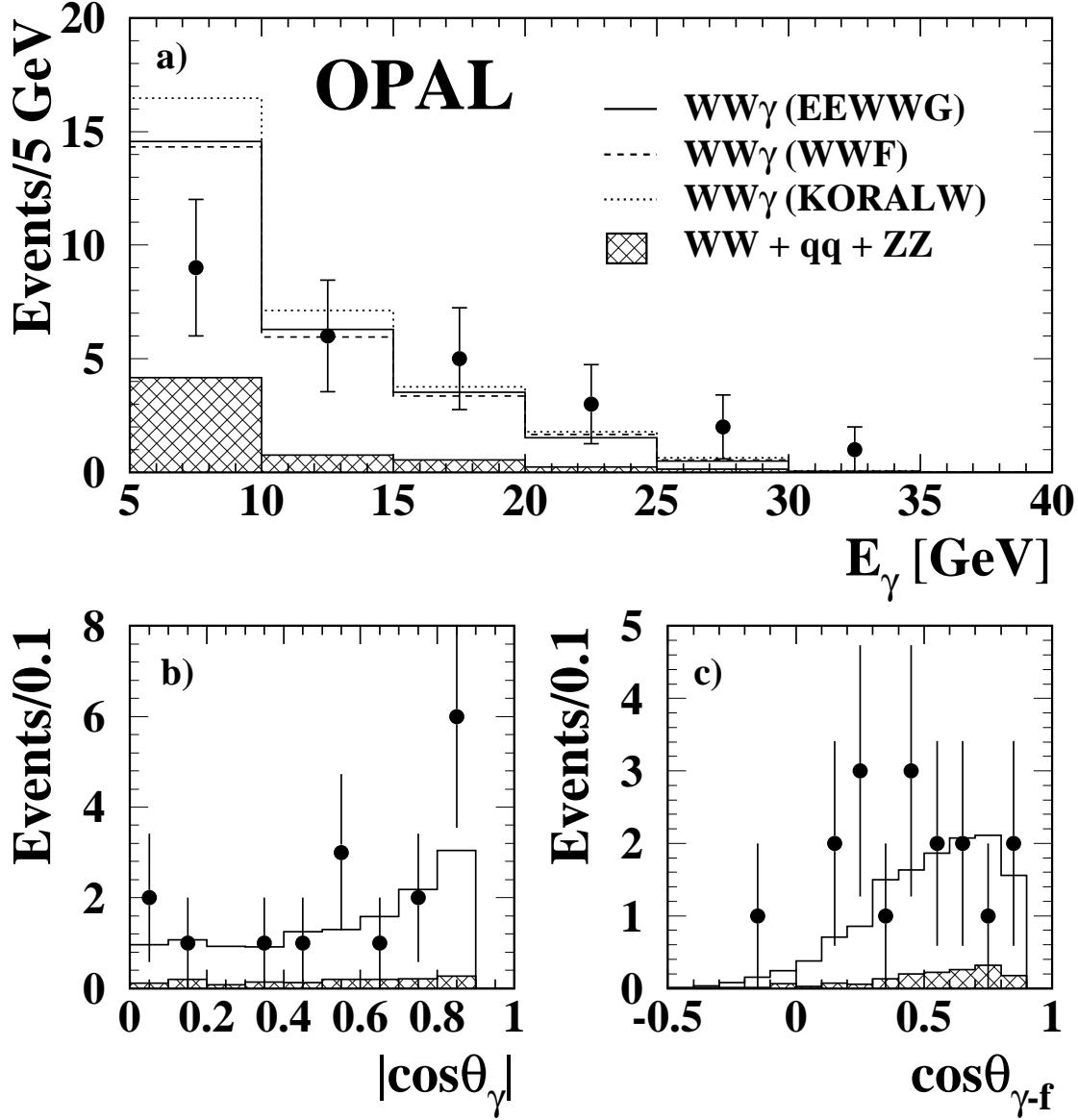}
 \caption{For selected $\WW\gamma$ events,
          a) shows the photon energy spectrum, b) 
          the cosine of the polar angle of the
          photon, and c) the cosine of the angle between the photon and 
          the nearest fermion. In b) and c) the cut $\Egam>10$~GeV has been
          applied.
          The 189~GeV data are shown by the points
          with error bars and the total SM Monte Carlo expectation shown 
          by the histogram. The hatched histograms indicate the contribution
          from non-\WW\ background and from fake photons in selected 
          \WW\ events. In Figure a) the predictions from \KoralW\ are
          compared to those from \WWF\ and \eewwg. The prediction of
          \eewwg\ also includes the estimated contribution from FSR 
          obtained using the \KoralW\ generator.}
 \label{fig:result_eg}
\end{figure}

Using the selection criteria defined in the previous section,
17 $\WWg$ events with $\Egam>10$~GeV are selected compared to the 
Monte Carlo expectation\footnote{Here \KoralW\ was used for the
expectation from $\WW(\gamma)$ and \Pythia\ was used to simulate the
backgrounds from \qq, \ZZ and \Wev} of 13.2 events. 
Figure~\ref{fig:result_eg}a shows the photon 
energy spectrum for the selected
$\WW\gamma$ event sample at $\roots=189$~GeV. Also shown are the data for
photon energies in the range $5-10$~GeV.
One event is observed with $\Egam>30$~GeV, 
compared to the expectation of 0.1. 
This event occurs in the electro-magnetic calorimeter barrel/endcap 
overlap region where the energy resolution is relatively poor.
Figure~\ref{fig:result_eg}b shows the distribution
of \cosg\ for the events with $\Egam>10$~GeV and
Figure~\ref{fig:result_eg}c shows the distribution of the cosine of the
angle between the photon and nearest charged fermion in the event. 
Good agreement between data and Monte Carlo is observed. 

\begin{figure}[htbp]
 \begin{center}
 \epsffile{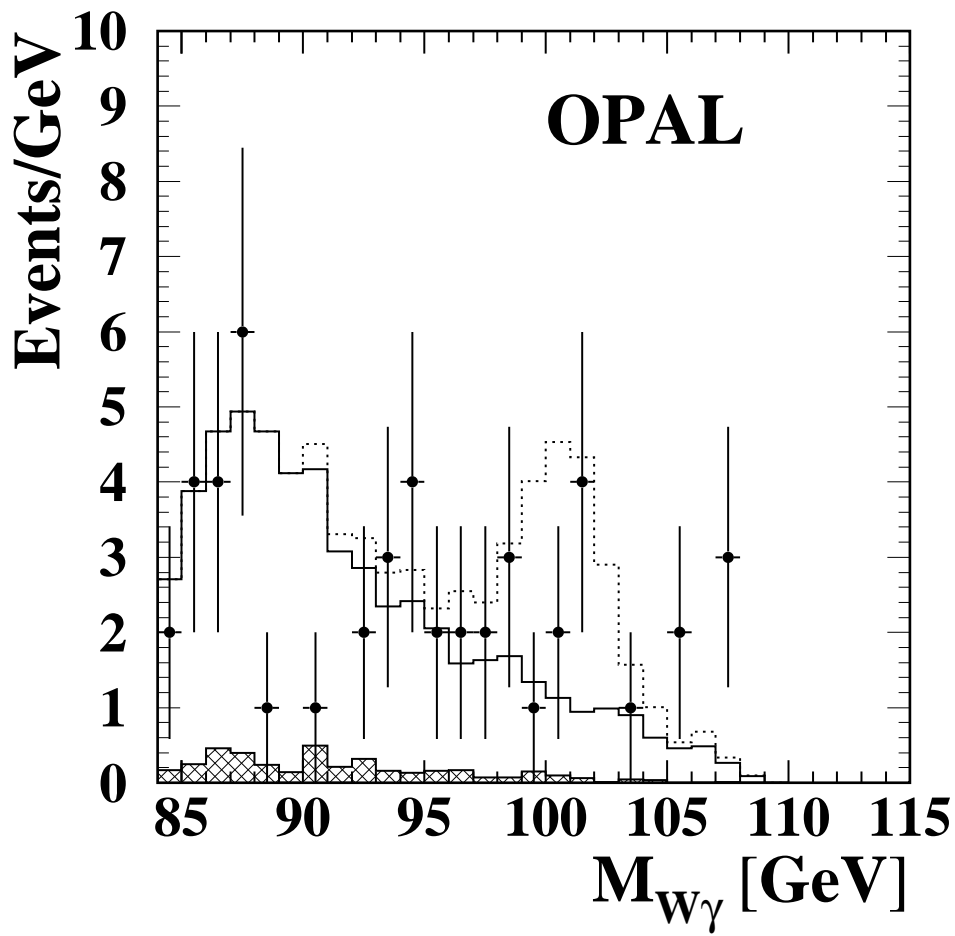}
 \caption{Reconstructed invariant mass of $\mathrm{W}^\pm\gamma$ 
          in selected \WWg\ events with $\Egam>5$~GeV (two entries per
          event). The 189~GeV data are shown by the points, the Standard
          Model expectation, determined from \KoralW\ and \Pythia\ is
          shown by the histogram. The shaded histogram shows the 
          non-$\WW(\gamma)$ background. Also shown, by the dotted
          histogram, is the expected  
          reconstructed mass 
          distribution  (arbitrary normalisation) 
          for a narrow resonance of mass 100~GeV
          which decays to $\mathrm{W}\gamma$.}
 \label{fig:wstar}
 \end{center}
\end{figure}

New physics could appear as resonant structure in the  
$\mathrm{W}\gamma$ invariant mass distribution. To investigate
this possibility, 
the invariant masses of the two $\mathrm{W}^\pm\gamma$ combinations
in selected \WWg\ events are obtained from an additional 
kinematic fit. The fit uses the constraints of energy and 
momentum conservation and the constraint that the invariant
masses of the reconstructed, 
$\fafb$ and $\fcfd$ systems are both equal to the W-mass.
%%($\Mw=80.33$~GeV). 
Only events for which the kinematic fit converges are retained. 
The $\mathrm{W}\gamma$ invariant mass is calculated
from the four-momenta of the four fermions and the
photon returned by the fit.  
Figure~\ref{fig:wstar} shows the reconstructed invariant mass 
distribution for the two $\mathrm{W}^\pm\gamma$ combinations for  
selected \WWg\ events with $\Egam>5$~GeV. No resonant structure is observed.
 
The $\WWg$ cross-section
for $\Egam>10$~GeV is determined within the acceptance defined
in Section~\ref{sec:sigdef}. 
The selection efficiency for events within this acceptance,
\effWWg, is evaluated using the \WWF\ Monte Carlo sample. 
Backgrounds from $\qq(\gamma)$ and $\ZZ$ are estimated using \Pythia. 
The background from $\WW$ events with a fake photon from the electro-magnetic
decays of mesons is estimated using \KoralW. These three sources 
are summed to form the total non-\WWg\ background, \sigback. 
Due to the steeply falling 
photon energy spectrum, the effect of the finite photon energy 
resolution is that more events from lower energies are reconstructed with
$\Egam>10$~GeV than vice versa. A correction, \cres, is applied to 
account for this migration. 
In addition, it is necessary to account for accepted
\WWg\ events from outside the invariant mass region used
to define the cross-section using a correction factor, \PWWg, which is
defined as the fraction of accepted $\fafb\fcfd\gamma$ events which have 
$\Mfafb > 73$~GeV and $\Mfcfd > 73$~GeV. 
The $\WWg$ cross-section is calculated from
\begin{eqnarray*}
     \sigWWg & = & {{ \PWWg (N_{\mathrm{obs}}- \sigback{\cal{L}})}
                             \over{ \effWWg\cres{\cal{L}}   }  },
 \label{eqn:sigWWg}
\end{eqnarray*}
where $N_{\mathrm{obs}}$  is the accepted number of events with $\Egam>10$~GeV
and ${\cal{L}}$ is the integrated luminosity.
The values of the quantities used to determine the cross-section
and associated uncertainties are listed in Table \ref{tab:xsecvals},
giving the result,
\begin{eqnarray*}
     \sigWWg & = & 136 \pm  37 \pm 8 \ \ \mathrm{fb},
\end{eqnarray*}
where the first error is statistical and the second systematic.
This result is consistent with the expectations ($85-102$~fb) of 
the \WWF, \KORALW\ and \EEWWG\ presented in Table~\ref{tab:smxsec}.
The sources of systematic uncertainty, summarised in Table
\ref{tab:systematics}, are described below.

\subsection{Systematic Uncertainties}

\bigskip
\noindent
\underline{\bf\boldmath \WW\ Selection Efficiency}:
The systematic uncertainty on the selection of \WW\ events is 
estimated to be 0.9\%. This estimate was obtained in a manner similar
to that described in \cite{bib:xs183}. The largest uncertainties
are related to the QCD and fragmentation modelling of jets.

\bigskip
\noindent
\underline{\bf\boldmath Photon Identification}:
A systematic uncertainty of 1\% is assigned to cover the
uncertainties in the simulation of the photon conversion rate and the 
accuracy of the simulation of the electromagnetic cluster shape. 

\bigskip
\noindent
\underline{\bf\boldmath Photon Isolation}:
The systematic error associated with the isolation requirements
depends on the accuracy of the Monte Carlo simulation of the fragmentation
process in hadronic jets. This is verified in $\Zzero\rightarrow\qq$ events
recorded at $\roots=91$~GeV during the 1998 run. For each selected
event, the inefficiency of the isolation requirements 
is determined for random orientations of the isolation cone and 
parametrised as a function of the angle between the
cone and the nearest jet. For all jet-cone angles the inefficiency
in the Monte Carlo and data agree to better than 1\%, consequently
a 1\% systematic error is assigned.

\bigskip
\noindent
\underline{\bf\boldmath Kinematic Fits}:
The \WWg\ event selections require that a kinematic fit converges 
and has a reasonable probability. Possible  mis-modelling of the 
detector response/resolution could result in a difference in the rates
at which the fits fail for data and Monte Carlo. This was checked by 
applying the kinematic fits used in the W mass analysis to all selected
\WW\ events and comparing the failure rates for data and Monte Carlo.
The difference in efficiency, $1.2\pm0.8\%$, is used as an estimate
the systematic error associated with the kinematic fits.

\bigskip
\noindent
\underline{\bf\boldmath ECAL energy scale and resolution}:
The systematic error (2.5\%) on the correction factor,
\cres, which accounts for feed-through from lower
energy photons, is 
estimated by varying the electro-magnetic calorimeter energy resolution by 
$\pm25\%$ and by varying the electro-magnetic calorimeter scale by $\pm1\%$. 
These variations were obtained by studying photons in 
$\epem\rightarrow\gamma\gamma$ and $\epem\rightarrow\epem\gamma$ 
at $\roots\sim\Mz$ and $\roots=189$~GeV, and by studying the energy
response to electrons. 

\bigskip
\noindent
\underline{\bf\boldmath Modelling of photons from jets}:
The modelling of photon candidates associated with the 
hadronic jets (both from
FSR and from $\pi^0$ and $\eta$ decays)  is studied by 
comparing the rate at which photons are identified in
$\Zzero\rightarrow\qq$ events from the 1998 calibration data
(\roots=91.2 GeV) to the \Pythia\  prediction.  
For 10~GeV$<\Egam<20$~GeV, there are 40\% more photon candidates
identified in the 
data than expected from the Monte Carlo. Above 20~GeV the data are consistent
with the Monte Carlo expectation.
The ratio of data to Monte Carlo is used to estimate an 
energy-dependent correction to the Monte Carlo 
expectation for the background from 
\WW\ events with fake photons. The size of
this correction is assigned as a systematic uncertainty. 

\bigskip
\noindent
\underline{\bf\boldmath ISR Modelling in $\qq\gamma$ Background}:
The dominant source of non-\WW\ background is from  
$\epem\rightarrow\Zgamma\rightarrow\qq\gamma$ where the identified photon
is  a genuine photon from ISR.  The accuracy of the simulation of 
initial state radiation with $|\cosg|<0.9$ is studied
using identified ISR photons in multi-hadronic 
events~\cite{bib:LEP2MH} recorded at 
$\roots=189$~GeV. A kinematic fit is used to reject events with additional 
radiation along the beam direction. In the range of photon energies of 
relevance to the \WWg\ analysis, $5~\mathrm{GeV}<\Egam<50$~GeV, 
the fractions of events with an identified photon in
data and Monte Carlo agree to better than 20\% which is assigned as a 
systematic uncertainty.

\bigskip
\noindent
\underline{\bf\boldmath Four-fermion Background}:
For the \WWg\ cross-section measurement \Pythia\ is used to estimate
the background from $\ZZ$ and single W production ($\Wev$).
This four-fermion background estimate was compared to that 
obtained using \grcff\ and \Excalibur. The largest difference is 
used to assign a systematic error.

\bigskip
\noindent
\underline{\bf\boldmath \WWg\ events from outside the signal definition}:
In order to account for accepted \WWg\ events from outside the
cross-section signal definition, 
a correction factor is used.
This factor, \PWWg, is determined to be 
$(85.3\pm0.7)\%$ from \WWF\ and $(85.0\pm1.0)$\% from \KoralW.
The $\WWF$ value is taken as the central value.
The comparison of these generators is sensitive to uncertainties in 
the treatment of FSR. Although no discrepancy is observed, 
the statistical error on the 
difference is taken as a systematic error (1.2\%). 
In addition, a systematic uncertainty of
1.5\% is assigned to account for the differences between
data and Monte Carlo in the modelling of photons from jets, described 
previously.

\section{Limits on Anomalous Quartic Couplings}

The selected $\WW\gamma$ events are used to set limits on possible
anomalous contributions to the  $\WW\gamma\gamma$ and 
$\WW\Zzero\gamma$ quartic gauge couplings. The limits are extracted 
from the measured differential cross-section 
as a function of the photon energy and photon polar angle.
 
There is currently no Monte Carlo available that implements 
anomalous quartic couplings and has contributions from all Standard
Model diagrams. The expectated differential \WWg\ cross-sections are
obtained using the \eewwgammamc\ program which allows for 
anomalous quartic couplings. Since FSR is not implemented in 
\eewwgammamc\ it is included as an additional background process. 
The Monte Carlo treatment of signal and background is described below.

\subsection{\boldmath \WWg\ Signal}

The \eewwgammamc\ calculation is used to determine the expected 
contribution from the ISR, WR, Standard Model QGC and
anomalous QGC diagrams within the signal definition of 
Section~\ref{sec:sigdef}. The only detector effect that is
included is a Gaussian smearing associated with 
the energy resolution for the photon. No smearing was applied
to the photon or fermion angles since these are relatively well measured.
The selection efficiency
is estimated (for events within the signal acceptance cuts) 
from the \WWF\ Monte Carlo and is 
parametrised as a function of photon energy.
The small dependence of the selection efficiency on \cosg\ is 
neglected in the determination of the limits on anomalous couplings.
The validity of this procedure is verified using a small sample of 
fully-simulated Monte Carlo events generated using the \EEWWG\ calculation.  
%%\footnote{this  
%%leads to a  conservative estimation of limits,
%%since anomalous contributions produce photons which are more central in the 
%%detector than the ISR spectrum, {\em i.e.} where the efficiency is higher.}

The contribution to the cross-section from any anomalous coupling 
increases rapidly with centre-of-mass 
energy due to the additional phase space 
for higher energy photons. Thus the contribution to the 
\WWg\ cross-section from anomalous QGCs is sensitive to higher-order QED 
radiative corrections. 
The \eewwg\ calculation currently only includes \WWg\ final states. For the
case of the QGC diagram, it therefore neglects the effect of additional
ISR which reduces the average effective centre-of-mass energy, 
$\rootsprime$. The effect of
collinear ISR from both the electron and positron has been incorporated
into the \eewwg\ calculation using the collinear radiator function from 
\Excalibur.  At $\roots=189$~GeV,
QED radiative corrections reduce the cross-section from
anomalous QGCs by 26\%.       

\subsection{Background Treatment}

The contribution from FSR is estimated using \KoralW, 
with FSR photons being  identified using the Monte Carlo generator 
information. \KoralW\ is also used to estimate the contribution from
other diagrams producing final states outside the signal definition of
Section~\ref{sec:sigdef}. Neglecting the contribution of anomalous
couplings to events outside the signal definition not only results in 
conservative limits but reduces the possible effects of interference with
Standard Model FSR diagrams. 
Background from \WW\ with a fake photon is calculated using the
corrected background from \KoralW. Similarly, the four-fermion and
\qq\ backgrounds are estimated using \Pythia. 
Table~\ref{tab:anombackground} summarises the background contributions
used in the anomalous coupling study. In the table, the \WWg\ 
background has been divided into that from ISR 
(outside the signal definition mass and/or acceptance cuts) and
that from FSR.          
 
\subsection{Results}

\begin{figure}[htbp]
 \epsfxsize=\textwidth
 \epsffile{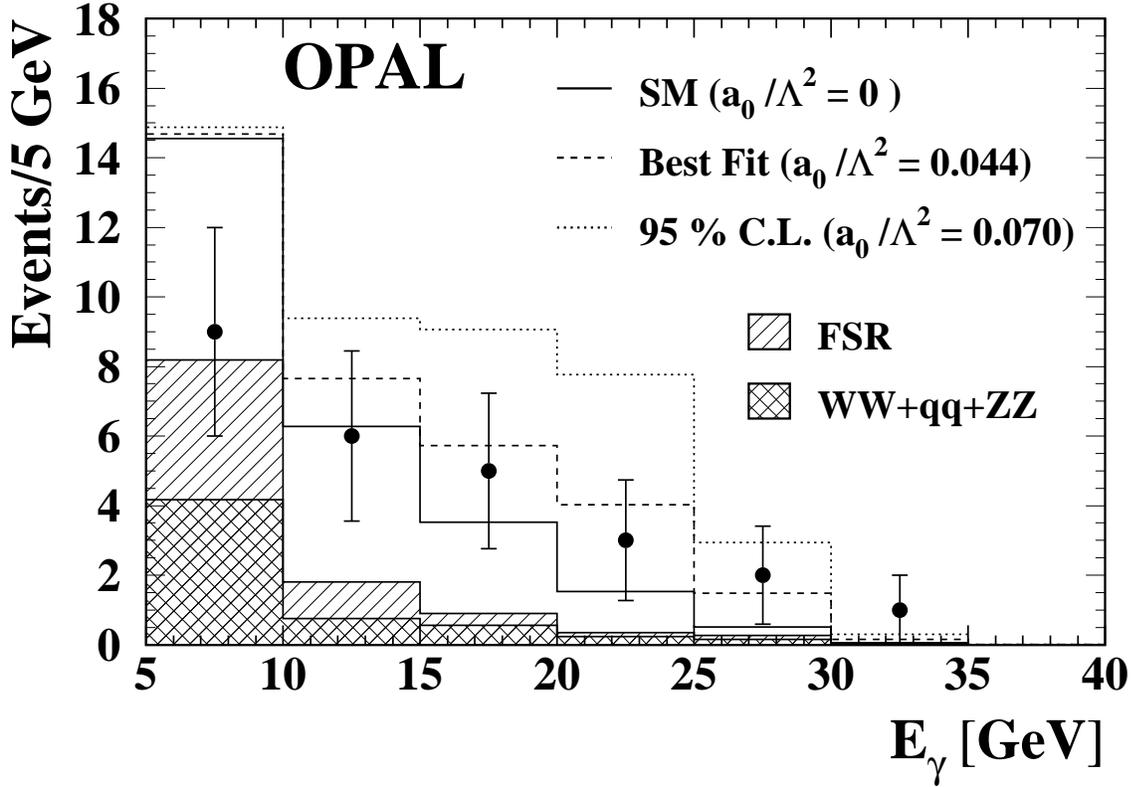}
 \caption{Energy spectrum of photons in candidate $\WWg$ events.
  The data, shown by the points with error bars, are compared to
  the expected distributions for three different values of 
  the anomalous coupling $a_0/\Lambda^2$;
  the Standard Model value $a_0/\Lambda^2=0$, the best fit value 
  $a_0/\Lambda^2=0.044$
  and the distribution corresponding to the 95\% confidence limit
  value of $a_0/\Lambda^2=0.070$. 
  The double hatched histogram shows the background
  from non-$\WW$ events and from $\WW$ events with a fake photon.
  The singly hatched histogram shows the contribution from final state
  radiation.}
 \label{fig:a0fit}
\end{figure}

For the case of Standard Model quartic gauge couplings, the 
expected photon energy spectrum for selected events estimated 
using the above procedure is shown in Figure \ref{fig:result_eg}a. 
The spectrum is in reasonable agreement with that 
obtained from the \WWF\ generator. 
By combining the predicted cross-sections from \eewwg\ and 
the backgrounds from Table~\ref{tab:anombackground} the expected 
distributions corresponding to non-zero anomalous coupling parameters can 
be generated. Figure~\ref{fig:a0fit} shows the expected photon energy
distributions for two values of the anomalous coupling $a_0/ \Lambda^2$
(in addition to the Standard Model prediction).

To set limits on possible anomalous couplings a binned maximum 
likelihood fit to the observed distribution of [\Egam, \cosg] is 
performed using bins of [5~GeV, 0.1], with the fit range restricted to 
$\Egam>10$~GeV.
The 95\% confidence level upper limits on the anomalous couplings
are obtained from the resulting likelihood curves:
\begin{eqnarray*}
   -0.070~\mathrm{GeV}^{-2} < &a_0/ \Lambda^2 &  < 0.070~\mathrm{GeV}^{-2}, \\ 
   -0.13~\mathrm{GeV}^{-2}  < &a_c/ \Lambda^2 &  < 0.19~\mathrm{GeV}^{-2}, \\
   -0.61~\mathrm{GeV}^{-2}  < &a_n/ \Lambda^2 &  < 0.57~\mathrm{GeV}^{-2}.  
\end{eqnarray*}
The expected 95\% confidence level upper limit on $a_0/ \Lambda^2$ is 0.045.
These limits from the data are higher than the expected limits 
due to the slight excess of high energy photons in the data.
The probability of obtaining a limit greater than or equal to the
observed limit is approximately 2\%. 
The limits were derived including the systematic uncertainties
described in the previous section. For these results, 
a 15\% theoretical uncertainty on the cross-section normalisation of 
the \eewwg\ calculation is assumed.

These are the first direct limits on anomalous quartic couplings.
However, indirect limits can be derived from radiative corrections
to LEP/SLD $\Zzero$ data. 
The experimental limits on deviations of
the S,U,T parameters~\cite{bib:SUT} (or equivalently  
$\epsilon_1,\epsilon_2,\epsilon_3$~\cite{bib:epsilons}) 
from their Standard Model values has been used to place constraints on
the anomalous $\WW\gamma\gamma$ and $\ZZ\gamma\gamma$
couplings~\cite{bib:Eboli}: $-0.0007 <a_0/\Lambda^2< 0.0001$ and 
$-0.0017 < a_c/\Lambda^2 < 0.0009$. However, these indirect limits are
obtained under a restrictive set of assumptions and their validity has 
been questioned~\cite{bib:lep2boudjema}. There are currently no limits 
on $a_n$ corresponding to an anomalous $\WW\Zzero\gamma$ coupling.

\section{Conclusions}

The first study of \WWg\ events 
produced in \epem\ collisions is presented. From the seventeen selected
events with $\Egam>10$~GeV, the \WWg\ production cross-section is
measured to be:
\begin{eqnarray*}
  \sigWWg = 136 \pm 37 \pm 8~\mathrm{fb},
\end{eqnarray*}  
within the $\fafb\fcfd\gamma$ 
generator level acceptance defined by $|\cosg| < 0.9$,
$\cosgf< 0.9$ and $\Mfafb,\Mfcfd > 73$~GeV, in agreement with Standard
Model expectation.

The distribution of the photon energy and polar angle is used to set 
limits on possible anomalous contributions to the
$\WW\gamma\gamma$ and $\WW\gamma\Zzero$ vertices:
\begin{eqnarray*}
   -0.070~\mathrm{GeV}^{-2} < & a_0/ \Lambda^2 & < 0.070~\mathrm{GeV}^{-2}, \\ 
    -0.13~\mathrm{GeV}^{-2} < & a_c/ \Lambda^2 & < 0.19~\mathrm{GeV}^{-2}, \\
    -0.61~\mathrm{GeV}^{-2} < & a_n/ \Lambda^2 & < 0.57~\mathrm{GeV}^{-2},  
\end{eqnarray*}
where $\Lambda$ represents the energy scale for new physics.
These are the first direct limits on anomalous quartic couplings.

\section{Acknowledgements}

We would like to thank James Stirling and Anja Werthenbach for providing 
the program \eewwg\ which is used to determine the effects of anomalous 
couplings in \WWg\ events. We also greatly appreciate their many 
useful suggestions and comments. We would also like to thank Markus Roth 
and Doreen Wackeroth and for providing the $4f\gamma$ cross-section 
calculations using the \Racoon\ program. We thank Fawzi Boudjema for his 
comments on this paper and for useful discussions.   

We particularly wish to thank the SL Division for the efficient operation
of the LEP accelerator at all energies
 and for their continuing close cooperation with
our experimental group.  We thank our colleagues from CEA, DAPNIA/SPP,
CE-Saclay for their efforts over the years on the time-of-flight and trigger
systems which we continue to use.  In addition to the support staff at our own
institutions we are pleased to acknowledge the  \\
Department of Energy, USA, \\
National Science Foundation, USA, \\
Particle Physics and Astronomy Research Council, UK, \\
Natural Sciences and Engineering Research Council, Canada, \\
Israel Science Foundation, administered by the Israel
Academy of Science and Humanities, \\
Minerva Gesellschaft, \\
Benoziyo Center for High Energy Physics,\\
Japanese Ministry of Education, Science and Culture (the
Monbusho) and a grant under the Monbusho International
Science Research Program,\\
Japanese Society for the Promotion of Science (JSPS),\\
German Israeli Bi-national Science Foundation (GIF), \\
Bundesministerium f\"ur Bildung, Wissenschaft,
Forschung und Technologie, Germany, \\
National Research Council of Canada, \\
Research Corporation, USA,\\
Hungarian Foundation for Scientific Research, OTKA T-029328, 
T023793 and OTKA F-023259.\\

\newpage

\begin{table}[htbp]
 \begin{center}
  \begin{tabular}{|l|r|c|r|}\hline
   Program             & Collinear ISR  & $\sigWWg$   \\ \hline
   \KoralW\            &  yes & $102.4 \pm 2.2$~fb  \\ 
   \eewwgammamc\       &  yes & $85.5  \pm 0.4$~fb  \\ 
   \WWF\               &  yes & $89.3  \pm 0.6$~fb  \\ \hline
   \eewwgammamc\       &   no & $106.1 \pm 0.4$~fb  \\ 
   \WWF\               &   no & $110.3 \pm 0.6$~fb  \\ 
   \Racoon\            &   no & $109.33\pm 0.02$~fb \\ \hline

\end{tabular}
 \caption{Standard Model cross-sections for the process 
     $\epem\rightarrow\WW\gamma$ within the 
    $\fafb\fcfd\gamma$ 
    generator level acceptance defined by $\Egam>10$~GeV, $|\cosg| < 0.9$,
    $\cosgf< 0.9$ and $\Mfafb,\Mfcfd > 73$~GeV. Where possible, values are
    quoted corresponding to the cases with and without collinear ISR.}
 \label{tab:smxsec}
\end{center}
\end{table}

\begin{table}[htbp]
 \begin{center}
  \begin{tabular}{|l|r|}
  \hline
   Quantity            &       Value                      \\ \hline\hline
   $N_{\mathrm{obs}}$  &             17                   \\ \hline
   $\mathcal{L}$       &     $183.1 \pm 0.6$~pb$^{-1}$    \\ \hline   
   \effWWg             &     $48.7  \pm 1.8\%$           \\ \hline 
   \cres               &     $1.065 \pm 0.026$            \\ \hline
   \sigback            &     $9.9   \pm 2.0~ \mathrm{fb}$ \\ \hline 
   \PWWg               &     $85.3  \pm 1.9\%$            \\ \hline
\end{tabular}
 \caption{Values of the quantities used in the determination of the
          \WWg\ cross-section. The errors include components from Monte Carlo
                 statistics and systematic uncertainties.}
 \label{tab:xsecvals}
\end{center}
\end{table}

\begin{table}[htbp]
\renewcommand{\arraystretch}{1.0}
\begin{center}
\begin{tabular}{|l|c|} \hline
Source                  &  $\Delta\sigWWg$\\ \hline\hline
$\WW$ Event Selection   &                      3.0~fb    \\ \hline
\qqg\ efficiency        &                      0.5~fb    \\ \hline
Photon Identification   &                      1.6~fb    \\ \hline
Photon Isolation        &                      1.6~fb    \\ \hline
Kinematic Fits          &                      1.6~fb    \\ \hline 
ECAL energy scale       &                      1.8~fb    \\ \hline
ECAL energy resolution  &                      2.8~fb    \\ \hline
Photons from jets       &                      2.7~fb    \\ \hline
ISR modelling \qqg      &                      1.9~fb    \\ \hline     
Acceptance Cuts         &                      2.8~fb    \\ \hline 
Four-fermion Events     &                      2.0~fb    \\ \hline
\PWWg                   &                      3.0~fb    \\ \hline\hline 
Total                   &                      7.7~fb    \\ \hline
\end{tabular}
\end{center}
\caption{Contributions to the systematic error on $\sigWWg$.}
\label{tab:systematics}
\end{table}

\begin{table}[htbp]
\renewcommand{\arraystretch}{1.0}
\begin{center}
\begin{tabular}{|l||c|c|c||c|c|c|} \hline
\Egam\ Range & \multicolumn{3}{c||}{non-\WWg\ Background} & 
                   \multicolumn{3}{c|}{\WWg\ Background}\\ \hline
          & \WW        & $4f$           & \qq   
          &  ISR       & FSR(lepton)    & FSR(quark)    \\ \hline\hline
$10-15$~GeV & $0.36\pm0.08$& $0.08\pm0.03$  & $0.32\pm0.08$ &
              $0.29\pm0.05$& $1.66\pm0.30$  & $0.74\pm0.14$ \\ \hline
$15-20$~GeV & $0.19\pm0.06$& $0.10\pm0.04$  & $0.26\pm0.07$ & 
              $0.25\pm0.05$& $0.15\pm0.04$  & $0.36\pm0.06$ \\ \hline
$20-25$~GeV & $0.05\pm0.03$& $0.02\pm0.01$  & $0.16\pm0.05$ &
              $0.06\pm0.03$& $0.01\pm0.01$  & $0.01\pm0.01$ \\ \hline 
$25-30$~GeV & $0.04\pm0.02$& $0.04\pm0.02$  & $0.07\pm0.03$ &
              $0.00\pm0.01$& $-$            & $0.11\pm0.04$ \\ \hline
$30-35$~GeV & $0.00\pm0.01$& $-$            & $0.02\pm0.02$ &
              $-$          & $-$            & $0.01\pm0.01$ \\ \hline\hline
%%$>10$~GeV  &  $0.65\pm$     & $-$            &     $-$     & 
%%                   $-$     & $-$            &     $-$  \\ \hline\hline
\end{tabular}
\end{center}
\caption{Expected numbers of background events used in the
         anomalous coupling analysis.
         The $\WW$ background consists of events with a  photon 
         from the decays of a $\pi^0$, $\eta$ {\em etc}. 
         In addition, $\WWg$ events from FSR diagrams 
         are considered as background and estimated using \KoralW.
         Accepted \WWg\ events from outside the 
         invariant mass cuts of the cross-section signal definition are
         also treated as background (denoted ISR). The quoted errors 
         include systematic contributions.}
\label{tab:anombackground}
\end{table}

\newpage

\newpage
 
\begin{thebibliography}{99}

\bibitem{bib:xs183}  \opalabbiendi, \EPC{8}{1999}{191}.

\bibitem{bib:lepxs} ALEPH Collaboration, R.\ Barate \etal, 
         \PLB{453}{1999}{107}; \newline
    DELPHI Collaboration, P.\ Abreu \etal, \PLB{456}{1999}{310}; \newline 
    L3 Collaboration, M.\ Acciarri \etal, \PLB{436}{1998}{437}. 


\bibitem{bib:gentle}
 D.\ Bardin \etal, Nucl. Phys. B, 
 Proc. Suppl. {\bf 37B} (1994) 148; \newline
 D.\ Bardin \etal,
 \CPC{104}{1997}{161}.

\bibitem{bib:Boos}
    E.\ Boos, H.-J.\ He, W.\ Kilian, A.\ Pukhov, C.-P.\ Yuan 
    and P.M.\ Zerwas, ``Strongly Interacting Vector Bosons at
    TeV $\epmem$ Linear Colliders'', DESY-96-256, hep-ph/9708310.

\bibitem{bib:LHC} A.S.\ Belyaev, O.J.P.\ Eboli, M.C.\ Gonzalez-Garcia, 
     J.K.\ Mizukoshi, S.F.\ Novaes and I.\ Zacharov, 
    \PRD{59}{1999}{015022}.

\bibitem{bib:Eboli}
 O.J.P.\ Eboli, M.C.\ Gonzal\'ez-Garcia and S.F.\ Novaes,
 \NPB{411}{1994}{381}.

\bibitem{bib:Brunstein} 
A.\ Brunstein, O.J.P.\ Eboli and M.C. Gonzalez-Garcia,
\PLB{375}{1996}{233}.

%%\bibitem{bib:Godfrey}
%%    S.\  Godfrey, ``Quartic Gauge Boson Couplings'', OCIP/C-95-5,
%%     hep-ph/9505252.

\bibitem{bib:Hagiwara}
    K.\ Hagiwara, R.D.\ Peccei, D.\ Zeppenfeld and K.\ Hikasa,
    \NPB{282}{1987}{253}. 

\bibitem{bib:Berlanger}
    G.\  B\'elanger and F.\ Boudjema, Phys. Lett.  {\bf B288} (1992) 201.

\bibitem{bib:Stirling1}
    G.\ Abu Leil and W.J.\ Stirling, J. Phys.  {\bf G21} (1995) 517.

\bibitem{bib:Stirling2}
    W.J.\ Stirling and A.\ Werthenbach,  DTP-99-30, hep-ph/9903315,
    to be published in Eur. Phys. J. {\bf C}.

\bibitem{bib:leptgc}
ALEPH  Collaboration, R.\ Barate \etal, \PLB{422}{1998}{369}; \newline 
DELPHI Collaboration, P.\ Abreu  \etal, \PLB{459}{1999}{382}; \newline
L3  Collaboration, M.\ Acciarri \etal, \PLB{413}{1997}{176}.

\bibitem{bib:tevatgc} CDF Collaboration, F.\ Abe \etal, 
              \PRL{78}{1997}{4536}; \newline
              D0 Collaboration, B.\ Abbott \etal, \PRD{58}{1998}{051101}.

\bibitem{bib:lep2boudjema}
    G.\  B\'elanger, F.\ Boudjema, Y.\ Kurihara, D.\ Perret-Gallix and
    A.\ Semenov, ``Bosonic Quartic Couplings at LEP2'', KEK-CP-087, 
    LAPTH-744/99, hep-ph/9908254 (1999).

\bibitem{bib:oops}
    A.\ Werthenbach,  private communication.

\bibitem{bib:WWF}
    G.J.\ van Oldenborgh, P.J.\ Franzini and A.\ Borrelli,  
      Comp. Phys. Comm. {\bf 83} (1994) 14.

\bibitem{bib:KORALW} 
 Program KORALW V1.42, 
 M.\ Skrzypek \etal, \CPC{94}{1996}{216};\newline
 M.\ Skrzypek \etal, \PLB{372}{1996}{289}; \newline
 M.\ Skrzypek \etal, \CPC{119}{1999}{1}. 

\bibitem{bib:Aeppli}
    A.\ Aeppli and D.\ Wyler,  Phys. Lett.  {\bf B262} (1991) 125.



\bibitem{bib:Photos}
E.\ Barberio and Z.\ Was,
Comp. Phys. Comm. 79 (1994) 291.

\bibitem{bib:Jetset}
T.~Sj{\"o}strand, Comp. Phys. Comm. {\bf 39} (1986) 374; \\
T.~Sj{\"o}strand and M.\ Bengtsson, Comp. Phys. Comm. {\bf 43} (1987) 367.


\bibitem{bib:GOPAL}
     J. Allison et al., Nucl. Instr. and Meth.  {\bf A317} (1992) 47.

\bibitem{bib:Denner}
    A.\  Denner, S.\ Dittmaier, M.\ Roth  and D.\ Wackeroth, 
    ``Predictions for all processes $\epem\rightarrow$ 4 fermions + $\gamma$'',
    BI-TP 99/10, PSI-PR-99-12, hep-ph/9904472, to be published in  
    Nucl. Phys. B.

\bibitem{bib:Doreen}  M.\ Roth and D.\ Wackeroth , Private Communication.

\bibitem{bib:detector}
  \opalahmet, \NIMA{305}{1991}{275}; \\
  B.E.\ Anderson {\em et al.}, IEEE Transactions on Nuclear Science, 
  {\bf 41} (1994) 845; \\
  S.\ Anderson {\em et al.}, \NIMA{403}{1998}{326}.

\bibitem{bib:lumi} \opalabbiendi, CERN-EP/99-097, 
Submitted to Eur. Phys J. {\bf C}. 

%%\bibitem{bib:energy} LEP energy working group, Beam Energy Update for 1998,
%%                     private communication.

\bibitem{bib:pythia}
     T. Sj\"ostrand, Comp. Phys. Comm. {\bf 82}  (1994) 74.  


\bibitem{bib:Herwig}
    G.\ Marchesini et al., Comp. Phys. Comm. {\bf 67} (1992) 465.

\bibitem{bib:kk} 
    S.\ Jadach, B.F.\ Ward and Z.\ Was,
    \PLB{449}{1999}{97}.

\bibitem{bib:GRC4F} 
 J.\ Fujimoto \etal, \CPC{100}{1997}{128}.

\bibitem{bib:excalibur}
 F.A.\ Berends, R.\ Pittau and R.\ Kleiss, 
 \CPC{85}{1995}{437}.



%%\bibitem{bib:YFSWW} 
%% YFSWW3 Program, S.\ Jadach \etal, \PLB{417}{1998}{326},

 
%%\bibitem{bib:pn378} 
%%    ``\WW\ production in \epem\ collisions at \roots=189~GeV'',
%%      OPAL Physics Note PN378, (1999).

\bibitem{bib:Durham}
 N.\ Brown and W.J.\ Stirling, \PLB{252}{1990}{657}; \\
 S.\ Bethke, Z.\ Kunszt, D.\ Soper and W.J.\ Stirling, \NPB{370}{1992}{310}; 
\\
 S.\ Catani \etal, \PLB{269}{1991}{432}; \\
 N.\ Brown and W.J.\ Stirling, \ZPC{53}{1992}{629}.

\bibitem{bib:OPAL_HGG} \opalackerstaff, \PLB{437}{1998}{218}.

\bibitem{bib:mass172} \opalackerstaff, \EPC{1}{1998}{395}.


\bibitem{bib:LEP2MH}
  \opalackerstaff, \EPC{6}{1999}{1}.


%% \bibitem{bib:IDGCON}
%%     OPAL conversion finder IDGCON, implemented in the ID package.
     
%%\bibitem{bib:pn232} 
%%    ``A test of QCD shower models with photon radiation from quarks'',
%%      OPAL Physics Note PN232, (1996).


\bibitem{bib:SUT} M.E.\ Peskin and T.\ Takeuchi, \PRD{46}{1992}{381}.

\bibitem{bib:epsilons} G.\ Altarelli, R.\ Barbieri and F.\ Caravaglios,
 \PLB{349}{1995}{145}. 


%%\bibitem{bib:xs189}   \opalabbiendi, Paper in preparation.

%%\bibitem{bib:qqqqQCD} \opalalexander, \ZPC{69}{1996}{543}.

%%\bibitem{bib:qqisrproblem} \opalalexander, \ZPC{69}{1996}{543}.



\end{thebibliography}
\end{document}